\documentclass[pdflatex,sn-apa]{sn-jnl}

\usepackage{amsmath,amssymb}
\usepackage{booktabs}
\usepackage{array}
\usepackage{tabularx}
\usepackage{makecell}

\begin{document}

\title[Fragility of Markowitz Portfolios]{From Efficient Frontier to Fragile Frontier: A Global Sensitivity Analysis of Markowitz Portfolios}

\author*[1]{\fnm{Stefano} \sur{Pellegrino}}
\email{stefano.pellegrino@unina.it}

\author[2]{\fnm{Giulia} \sur{Vannucci}}
\email{giulia.vannucci@unina.it}

\author[2]{\fnm{Roberta} \sur{Siciliano}}
\email{roberta.siciliano@unina.it}

\affil*[1]{
\orgdiv{Department of Physics},
\orgname{University of Naples Federico II},
\orgaddress{
\street{Via Cintia 21},
\city{Naples},
\postcode{80125},
\country{Italy}
}}

\affil[2]{
\orgdiv{Department of Electrical Engineering and Information Technologies},
\orgname{University of Naples Federico II},
\orgaddress{
\street{Via Claudio 21},
\city{Naples},
\postcode{80125},
\country{Italy}
}}

\abstract{In mean-variance portfolio analysis, the efficient frontier represents the optimal trade-off between expected return and risk, assuming stable underlying parameters. This paper investigates portfolio fragility: the instability of optimal weights, risk-adjusted performance, and diversification when model inputs and construction choices are jointly perturbed. Combining constrained Markowitz optimization with variance-based global sensitivity analysis (Sobol indices), we map out how input uncertainty and portfolio-construction choices propagate along the target-return dimension. Using an empirical universe of multi-asset exchange-traded funds (ETFs), we find a distinct transition in the sensitivity structure: in the baseline experiment, lower target returns are dominated by $\ell_2$ regularization, whereas aggressive return requirements become increasingly sensitive to the weight cap and expected-return perturbations. This shift coincides with a sharp drop in effective diversification and a rise in weight dispersion. We extend the analysis to a multi-universe fragility atlas, showing that under a common absolute concentration rule, the smallest universe is weight-cap-driven in the aggressive return region, while larger sampled universes remain more often regularization-driven. The fragile frontier serves as a direct diagnostic tool to evaluate the structural robustness of constrained optimizers without altering the underlying allocation rule.}


\maketitle

\section{Introduction} 
Mean-variance portfolio optimization remains a central reference point for portfolio construction. Since \citet{markowitz1952,markowitz1959}, the efficient frontier has provided a tractable representation of the trade-off between expected return and variance. Yet this representation is silent on a different question: does the optimizer itself become less stable as the target return becomes more demanding, and which inputs drive that instability?

The instability of mean-variance optimization is well documented. Optimal weights can react strongly to small changes in expected returns and covariances, especially when expected returns are estimated with substantial noise \citep{michaud1989,best1991,green1992,chopra1993,brittenjones1999}. Statistically, the estimated optimal portfolio is a random object induced by uncertain inputs; economically, this may produce concentrated portfolios, large reallocations, and high turnover. A broad literature has therefore developed stabilizing devices, including Bayesian shrinkage of expected returns \citep{frost1986,jorion1986,black1992}, covariance shrinkage and factor-structured estimation \citep{ledoit2004well,ledoit2004honey,fan2011high,ledoit2017goldilocks}, norm penalties and portfolio constraints \citep{jagannathan2003risk,demiguel2009generalized,gotoh2011,brodie2009}, and robust optimization under parameter uncertainty \citep{goldfarb2003,fabozzi2010,esfahani2018}. Empirical evidence also shows that naive or highly stabilized allocations are often difficult to dominate out of sample \citep{demiguel2009optimal,tu2011,duchin2009,kan2022,kolm2014}.  

What remains less explored is a more specific diagnostic problem. Classical sensitivity analysis of constrained mean-variance portfolios has shown that optimal portfolios may change structure when parameters or constraint right-hand sides vary, often through changes in the active constraint set \citep{best1991sensitivity}. The present paper differs by using variance-based global sensitivity analysis (GSA), in which multiple inputs are perturbed jointly and their contributions to output uncertainty are decomposed into direct and interaction-driven components.  

Variance-based GSA is particularly useful in this application because it treats the constrained optimizer as a map from uncertain inputs to portfolio outputs and decomposes the resulting output variance into contributions associated with individual inputs and their interactions \citep{sobol2001,saltelli2002,saltelli2008,saltelli2010}. Modern sensitivity-analysis workflows also emphasize estimator choice, computational budget, convergence assessment, and careful interpretation of assumptions such as input independence \citep{iooss2015,pianosi2016,razavi2021}. In portfolio optimization, this makes GSA useful both as a parameter-ranking device and as a diagnostic tool for understanding how uncertainty propagates through a constrained optimizer.  

The paper contributes in three directions. First, it introduces the notion of a \emph{fragile frontier}, a diagnostic complement to the efficient frontier that describes how portfolio fragility evolves with the target return. Second, it applies established Sobol first-order and total-order sensitivity measures to constrained mean-variance optimization, separating direct from interaction-driven effects; the contribution lies in their use as diagnostics of the optimization map rather than in a new Sobol estimator. Third, the analysis is extended to a fragility atlas over relative target return and universe size, using repeated nested-stratified ETF baskets to examine the role of $N$ under a controlled basket-construction design.

The remainder of this paper is organized as follows. Section~\ref{sec:methodology} describes the constrained optimization model and the global sensitivity framework. Section~\ref{sec:data} presents the data and experimental design. Section~\ref{sec:results} details the empirical results and the multi-universe atlas, while Section~\ref{sec:discussion} discusses the implications and limitations of the findings, and Section~\ref{sec:conclusion} concludes. 

\section{Methodology}
\label{sec:methodology}

\subsection{Constrained Mean-Variance Optimization}
Let $w \in \mathbb{R}^N$ denote portfolio weights, $\mu \in \mathbb{R}^N$ expected returns, and $\Sigma \in \mathbb{R}^{N\times N}$ the covariance matrix of asset returns. For a target return $r^\star$, we consider the constrained long-only problem
\begin{equation}
\begin{aligned}
\min_w \quad & w^\top \Sigma w + \lambda \|w\|_2^2 \\
\text{s.t.} \quad & w^\top \mu \ge r^\star, \\
& \mathbf{1}^\top w = 1, \\
& w_i \ge 0, \quad i=1,\dots,N, \\
& w_i \le c, \quad i=1,\dots,N,
\end{aligned}
\label{eq:markowitz}
\end{equation}
where $\lambda\ge 0$ is an $\ell_2$ regularization parameter and $c$ is an upper bound on individual asset weights. The regularization term shrinks the solution toward less extreme allocations, while the weight cap directly controls concentration. In the sensitivity analysis that follows, $\lambda$ and $c$ are not treated as fixed calibration choices but as uncertain inputs, so that their stabilizing role can itself be diagnosed.  

\subsection{Variance-Based Global Sensitivity Analysis}
\label{subsec:gsa}

Variance-based global sensitivity analysis (GSA) studies how uncertainty in model inputs propagates to uncertainty in model outputs. In the present setting, the model is the constrained portfolio optimizer: for each target-return level and each realization of the uncertain inputs, the optimizer returns a set of portfolio-level quantities such as variance, Sharpe ratio, weights, and effective diversification. We use GSA to identify which modeling and construction inputs account for the variability of these outputs along the target-return dimension.

Let
\[
Y=f(X_1,\dots,X_k)
\]
denote a scalar model output and let \(X=(X_1,\dots,X_k)\) be the vector of uncertain inputs. Under the classical assumption of mutually independent inputs, the variance of \(Y\) can be decomposed into additive contributions associated with individual inputs and their interactions:
\[
\mathrm{Var}(Y)=\sum_{i=1}^k V_i+\sum_{i<j}V_{ij}+\cdots+V_{1\cdots k}.
\]
The term \(V_i\) measures the contribution of input \(X_i\) acting alone, whereas higher-order terms such as \(V_{ij}\) measure the contribution of interactions between inputs.

The first-order Sobol index is defined as
\[
S_i=\frac{V_i}{\mathrm{Var}(Y)}.
\]
It measures the fraction of output variance that can be attributed to input \(X_i\) alone. A large \(S_i\) therefore indicates that the input has a strong direct effect on the output. The total-order index is
\[
S_{T_i}=1-\frac{\mathrm{Var}\!\left(\mathbb{E}[Y\mid X_{\sim i}]\right)}{\mathrm{Var}(Y)},
\]
where \(X_{\sim i}\) denotes all inputs except \(X_i\). The total-order index measures the contribution of \(X_i\) including all interaction terms in which \(X_i\) participates. Thus, comparing \(S_i\) and \(S_{T_i}\) allows direct effects to be distinguished from interaction-driven sensitivity.

In practice, the Sobol indices are estimated through the Saltelli sampling scheme \citep{saltelli2002,saltelli2010}. Two base sample matrices,
\[
\mathbf{A},\mathbf{B}\in\mathbb{R}^{N_S\times k},
\]
are generated over the uncertain-input space, where \(N_S\) is the base sample size and \(k\) is the number of uncertain inputs. For each input \(X_i\), an additional hybrid matrix \(\mathbf{A}_{B}^{(i)}\in\mathbb{R}^{N_S\times k}\) is constructed by taking all columns from \(\mathbf{A}\), except for the \(i\)-th column, which is taken from \(\mathbf{B}\). Let
\[
Y_A^{(j)}=f\!\left(\mathbf{A}_{j\cdot}\right),
\qquad
Y_B^{(j)}=f\!\left(\mathbf{B}_{j\cdot}\right),
\qquad
Y_{A_B^{(i)}}^{(j)}
=
f\!\left(\mathbf{A}_{B,j\cdot}^{(i)}\right)
\]
denote the corresponding model evaluations for row \(j\). The first-order and total-order indices are estimated as
\[
\widehat{S}_i
=
\frac{
\displaystyle
\frac{1}{N_S}
\sum_{j=1}^{N_S}
Y_B^{(j)}
\left(
Y_{A_B^{(i)}}^{(j)}-Y_A^{(j)}
\right)
}{
\widehat{\operatorname{Var}}(Y)
},
\]
and
\[
\widehat{S}_{Ti}
=
\frac{
\displaystyle
\frac{1}{2N_S}
\sum_{j=1}^{N_S}
\left(
Y_A^{(j)}-Y_{A_B^{(i)}}^{(j)}
\right)^2
}{
\widehat{\operatorname{Var}}(Y)
},
\]
where \(\widehat{\operatorname{Var}}(Y)\) is estimated from the pooled outputs associated with \(\mathbf{A}\) and \(\mathbf{B}\). With second-order terms disabled, the design therefore requires \(N_S(k+2)\) model evaluations: \(N_S\) evaluations for each of the two base matrices and \(N_S\) evaluations for each of the \(k\) hybrid matrices.

Following the standard variance-based sensitivity-analysis framework, we use first-order and total-order indices as complementary diagnostics rather than as competing measures \citep{sobol2001,saltelli2002,saltelli2008,saltelli2010}. The first-order index identifies inputs with strong isolated effects, while the total-order index identifies inputs that remain important once interactions with the rest of the uncertainty space are included. This distinction is particularly relevant in constrained optimization, where the effect of a modeling input may depend on whether other constraints or regularization terms are active.

We use the difference
\[
G_i = S_{T_i}-S_i
\]
as a compact diagnostic of interaction-driven sensitivity. When \(G_i\) is small, the total influence of \(X_i\) is close to its first-order contribution, suggesting that most of its effect is direct. When \(G_i\) is large, the total contribution of \(X_i\) substantially exceeds its first-order effect, indicating that the input affects the output mainly through interactions with other uncertain inputs. The empirical classification of interaction gaps into low, moderate, and high categories is described in Section~\ref{subsec:sobol_empirical_design}.

\subsection{Fragility Diagnostics and Auxiliary Score}
\label{subsec:fragility_score}

Portfolio fragility is assessed through a set of output-instability diagnostics computed across the Sobol evaluations at each target-return level. For a portfolio with weights \(w\) under perturbed moments \(\tilde{\mu}\) and \(\tilde{\Sigma}\), portfolio variance and Sharpe ratio are given by
\[
\sigma_p^2 = w^\top \tilde{\Sigma} w,
\qquad
\mathrm{SR}
=
\frac{w^\top \tilde{\mu}-r_f}
{\sqrt{w^\top \tilde{\Sigma} w}},
\]
where \(r_f=0.02\). Effective diversification is measured by
\[
N_{\mathrm{eff}}
=
\frac{1}{\sum_{i=1}^{N} w_i^2}.
\]

The diagnostics include the standard deviation of portfolio variance, the standard deviation of the Sharpe ratio, the average standard deviation of portfolio weights, the standard deviation of a representative asset weight in the baseline case, and the mean effective number of assets. These quantities describe complementary aspects of allocation instability and diversification.

For visualization and onset detection, we also use the auxiliary standardized score
\begin{equation}
F_\star=
\frac{
z(D_\sigma)+z(D_S)+z(D_w)-z(\bar N_{\mathrm{eff}}/N)
}{4},
\label{eq:fstar}
\end{equation}
where \(D_\sigma\) is variance dispersion, \(D_S\) is Sharpe-ratio dispersion, \(D_w\) is average weight dispersion, and \(\bar N_{\mathrm{eff}}/N\) is the effective-diversification ratio. The negative sign reflects that lower relative diversification corresponds to greater fragility.

The components of \(F_\star\) are standardized using fixed reference distributions. In the baseline analysis, the dense and upper-tail-refined grids share a common reference because they are alternative discretizations of the same target-return experiment. In the atlas, the reference is the low-target regime pooled across replicates and universe sizes. Accordingly, \(F_\star\) measures relative departure from the relevant reference regime.

\subsection{Onset Detection}
\label{subsec:onset_detection}

The relative onset proxy is computed from \(F_\star\). For each target grid, the first third of target levels defines the low-target regime. We compute its median and interquartile range and define the cutoff as
\[
\mathrm{median}_{\mathrm{low}}(F_\star)+1.5\,\mathrm{IQR}_{\mathrm{low}}(F_\star).
\]
The onset proxy is the first target return at which \(F_\star\) exceeds this cutoff. We also report the largest jump in \(F_\star\) and the first target return at which the dominant driver shifts from \(\ell_2\) penalty to weight cap. These are diagnostic transition indicators, not universal thresholds.

\section{Data and Empirical Design}  
\label{sec:data}
The baseline study uses eight liquid ETFs: SPY, QQQ, EFA, IEF, TLT, GLD, VNQ, and LQD. These instruments provide a compact cross-asset universe covering U.S. equity, growth equity, international equity, government bonds, long-duration bonds, gold, real estate, and investment-grade credit. The atlas uses the larger 64-ETF master universe described in Appendix~\ref{app:master_universe}. The baseline composition is reported in Table~\ref{tab:baseline_etfs}.  

Daily adjusted prices were obtained from Yahoo Finance for the period January 2018--December 2025 and converted into log returns. Expected returns and covariances were estimated as annualized sample means and sample covariances. Standard sample estimators were used to keep the focus on uncertainty propagation through the constrained optimizer rather than on the properties of a particular moment estimator. Sharpe ratios were computed using a constant annual risk-free rate of 2\%, treated as a fixed benchmark for comparability across scenarios. All tickers used in the study were retrieved successfully. No synthetic data were used, and no ticker was excluded from the analysis.

\begin{table}[htbp]
\centering
\caption{Baseline ETF universe.}
\label{tab:baseline_etfs}
\small
\setlength{\tabcolsep}{5pt}
\begin{tabularx}{\textwidth}{l l X}
\toprule
Ticker & Asset class & Role in the baseline universe \\
\midrule
SPY & U.S. equity & Broad U.S. equity exposure \\
QQQ & U.S. growth equity & Technology/growth-oriented equity exposure \\
EFA & International equity & Developed ex-U.S. equity exposure \\
IEF & Treasury bonds & Intermediate-duration government bonds \\
TLT & Long Treasury bonds & Long-duration government bonds \\
GLD & Gold & Real-asset and defensive diversifier \\
VNQ & Real estate & U.S. REIT exposure \\
LQD & Corporate bonds & Investment-grade credit exposure \\
\bottomrule
\end{tabularx}
\normalsize
\end{table}

\subsection{Empirical Uncertainty Design}
\label{subsec:uncertainty_design}

For each nominal target-return level, the optimization map is perturbed through four uncertain inputs: the expected-return perturbation scale \(s_\mu\), the covariance perturbation scale \(s_\Sigma\), the weight cap \(c\), and the \(\ell_2\) regularization intensity \(\lambda\). The first two inputs are scalar amplitudes rather than asset-specific perturbations.

Let \(\hat{\mu}\) and \(\hat{\Sigma}\) denote the sample estimates of expected returns and covariances. To construct the perturbation templates, we generate, under a fixed random seed, a vector \(g_\mu\in\mathbb{R}^N\) and a matrix \(G_\Sigma\in\mathbb{R}^{N\times N}\) whose entries are drawn from a standard normal distribution. We then define
\[
d_\mu
=
\frac{g_\mu}{\lVert g_\mu\rVert_2},
\qquad
H_\Sigma
=
\frac{G_\Sigma+G_\Sigma^\top}{2},
\qquad
D_\Sigma
=
\frac{H_\Sigma}{\lVert H_\Sigma\rVert_F},
\]
where \(\lVert\cdot\rVert_2\) denotes the Euclidean norm and \(\lVert\cdot\rVert_F\) the Frobenius norm. Thus, \(d_\mu\) is a unit direction in the expected-return space, whereas \(D_\Sigma\) is a symmetric unit-Frobenius-norm direction in the covariance-matrix space. In the baseline experiment, a single pair of perturbation templates is generated and held fixed across both target grids, all target-return levels, and all Sobol evaluations. In the atlas, a separate pair is generated for each replicate and universe size and then held fixed across all relative target levels and Sobol evaluations within that configuration. Consequently, the uncertain inputs control only the amplitudes of the perturbations within each configuration, while the perturbation directions remain fixed. The nested basket construction therefore controls asset inclusion across universe sizes, whereas the perturbation templates are generated separately for each universe size. 

Expected returns are perturbed along \(d_\mu\), using asset volatilities as natural scales:
\[
\tilde{\mu}
=
\hat{\mu}
+
s_\mu
\left(
\sqrt{\operatorname{diag}(\hat{\Sigma})}
\odot d_\mu
\right),
\]
where \(s_\mu=\texttt{mu\_scale}\) and \(\odot\) denotes element-wise multiplication. The covariance matrix is perturbed along \(D_\Sigma\):
\[
\tilde{\Sigma}
=
\Pi_{\varepsilon}
\left(
\hat{\Sigma}
+
s_\Sigma\bar{\sigma}^2D_\Sigma
\right),
\]
where \(s_\Sigma=\texttt{sigma\_scale}\) and \(\bar{\sigma}^2\) is the average diagonal element of \(\hat{\Sigma}\). The operator \(\Pi_{\varepsilon}\) symmetrizes the perturbed matrix and floors its eigenvalues at \(\varepsilon=10^{-8}\), ensuring a positive-definite covariance matrix for the optimization. The analysis therefore considers two aggregate market-parameter stress channels rather than treating every mean and covariance element as a separate uncertain input.

Throughout the main analysis, \(\hat{\Sigma}\) denotes the annualized sample covariance matrix. 
To assess whether the results are sensitive to noise in this initial covariance estimate, Appendix~\ref{app:robustness} reports a covariance-filtering robustness check in which the full baseline analysis is repeated with a Random Matrix Theory (RMT)-filtered covariance matrix, while keeping \(\hat{\mu}\), the target grids, the Sobol input space, and the perturbation design unchanged.

The bounds in Table~\ref{tab:uncertain_inputs} define a controlled stress domain for the optimizer, rather than estimated probability intervals for market parameters. In the main Sobol design, each input is sampled uniformly over its corresponding range. The ranges are chosen to induce non-negligible variation in portfolio outputs while keeping the experiment within economically interpretable and numerically stable regions. Consequently, the reported Sobol indices are conditional on this perturbation domain.  For each grid level, \(r^\star\) denotes the nominal return requirement used to index the experiment. Perturbations in expected returns and the weight-cap parameter may shift the feasible return interval of an individual configuration. When the nominal target falls outside that interval, the optimization is performed at the corresponding feasible boundary. This preserves feasibility throughout the uncertainty design and allows the upper-tail analysis to capture the interaction between increasingly demanding return requirements and binding portfolio constraints. In the following, target return refers to the nominal grid level used to index the experiment.

\begin{table}[htbp]
\centering
\caption{Uncertain inputs in the Sobol design.}
\label{tab:uncertain_inputs}
\small
\setlength{\tabcolsep}{5pt}
\begin{tabularx}{\textwidth}{c c c X}
\toprule
Input & Baseline range & Atlas range & Interpretation \\
\midrule
\(s_\mu\) &
\([0.00,0.35]\) &
\([0.00,0.35]\) &
Expected-return perturbation scale. \\

\(s_\Sigma\) &
\([0.00,0.35]\) &
\([0.00,0.35]\) &
Covariance perturbation scale. \\

\(c\) &
\([0.20,0.60]\) &
\([0.27,0.60]\) &
Upper bound on individual asset weights. \\

\(\lambda\) &
\([0.00,0.08]\) &
\([0.00,0.08]\) &
Intensity of the \(\ell_2\) regularization term. \\
\bottomrule
\end{tabularx}
\normalsize
\end{table}

The lower bound on the weight cap \(c\) is feasibility-driven. In the baseline eight-asset universe, the lower bound is 0.20. In the atlas, the common lower bound is 0.27, corresponding to \(1/4+0.02\), because the smallest atlas universe has \(N=4\). The same absolute cap range is retained across all universe sizes, so that each basket is evaluated under a common concentration rule. The atlas therefore captures how a fixed concentration policy interacts with asset cardinality and the target-return requirement.

The main Sobol design is based on independent stress channels, which is the setting in which the classical Sobol variance decomposition admits its direct functional-ANOVA interpretation. In the present application, \(s_\mu\) and \(s_\Sigma\) represent distinct perturbation channels applied to the optimization map rather than jointly estimated econometric parameters. As a complementary robustness exercise, we couple the two scales through a Gaussian copula with \(\rho=0.5\), introducing controlled dependence between the two market-uncertainty channels \citep{Kucherenko2012}. The resulting input-output associations are summarized using absolute Pearson correlations at representative low, middle, and high target levels.

\subsection{Sobol Sampling and Gap Classification}
\label{subsec:sobol_empirical_design}

The Saltelli sampling design and the first-/total-order estimators described in Section~\ref{subsec:gsa} are implemented in Python using \texttt{SALib} \citep{Herman2017,Iwanaga2022}. Samples are generated using the Saltelli quasi-random design \citep{saltelli2002,saltelli2010}, which provides estimates of first- and total-order indices without explicitly estimating second-order terms. With \(k=4\) inputs and base sample size \(N_S=512\), the first/total-order design requires
\[
N_S(k+2)=3,072
\]
model evaluations per target-return level. Because the Sobol indices are estimated from a finite quasi-random sample, the interpretation focuses on persistent patterns across target regions, outputs, grid refinements, and robustness checks rather than on isolated point estimates. A sample-size check over \(N_S\in\{32,64,128,256,512\}\) is reported in Appendix~\ref{app:robustness}.
For compact comparison across target levels, we define the dominant driver as the input with the largest mean total-order Sobol index across the outputs considered in each experiment. The baseline summary uses portfolio variance, Sharpe ratio, SPY weight, and effective number of assets. In the atlas, the SPY-specific output is omitted because basket composition changes across universe sizes and replicates; the ranking is therefore based on portfolio variance, Sharpe ratio, and effective number of assets.

The interaction gaps \(G_i=S_{T_i}-S_i\) are classified relative to the empirical dense-grid baseline distribution. The dense-grid baseline distribution contains 400 observed gaps, with \(q_{25}=0.004395\), median \(=0.036315\), and \(q_{75}=0.095772\). A gap is classified as low if \(G_i\le q_{25}\), moderate if \(q_{25}<G_i<q_{75}\), and high if \(G_i\ge q_{75}\).

This classification is descriptive and sample-dependent because the reference quartiles are estimated from the same empirical experiment being interpreted. The labels are used only to distinguish relatively small, intermediate, and large interaction gaps within the dense-grid baseline experiment.

\subsection{Target Grids and Atlas Construction}
\label{subsec:target_grids_atlas}

The baseline analysis uses two target-return grids: a dense uniform grid and an upper-tail-refined grid. The refined grid is not a separate empirical experiment; it is a local discretization check designed to test whether the detected transition is a grid artifact.

The atlas extends the design to universe sizes \(N=4,8,12,24\). It is built from a 64-ETF master universe grouped into six economic strata. For each of 10 replicates, nested baskets are constructed so that \(\mathcal U_4\subset\mathcal U_8\subset\mathcal U_{12}\subset\mathcal U_{24}\). The 24-asset case is therefore a sampled sub-universe from the master universe rather than the full master universe itself. Target returns in the atlas are expressed on a relative scale within each basket's feasible interval.

\section{Results}
\label{sec:results}

\subsection{Baseline Allocation Paths and Weight Dispersion}
\begin{figure}[htbp]
    \centering
    \includegraphics[width=1\textwidth]{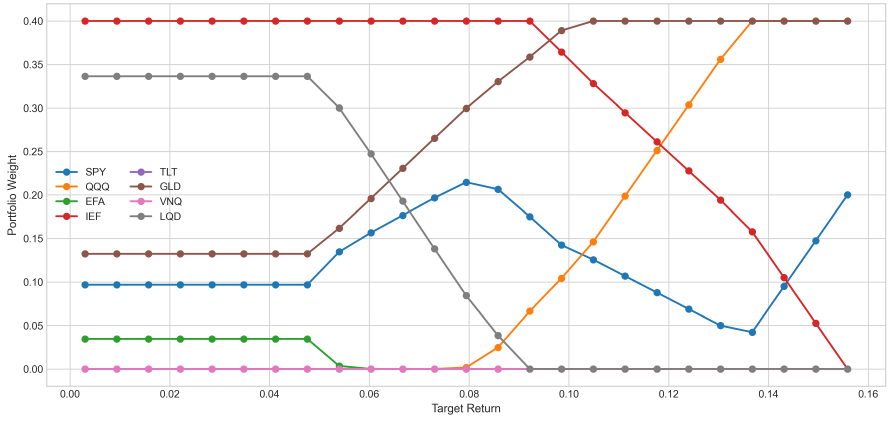}
    \caption{Baseline Markowitz weights as a function of target return}
    \label{fig:baselineweights}
\end{figure}

Figure~\ref{fig:baselineweights} shows how the baseline allocation changes along the target-return grid. At low target returns, the allocation is spread across defensive and diversifying assets, so the regularization term mainly governs how strongly the optimizer smooths the solution. As the target return increases, several assets either approach the weight cap or are driven close to zero, and the allocation becomes concentrated in a smaller set of return-contributing exposures. The same change in allocation structure also appears in the sensitivity results below, where concentration controls become increasingly important in the upper part of the frontier.

Table~\ref{tab:dense_summary} reports selected target configurations. Average weight dispersion rises from 0.029 at the minimum target to 0.066 at the maximum target, while the SPY weight dispersion rises from 0.008 to 0.137. Over the same range, mean effective diversification falls from 5.950 to 2.942. The standardized score $F_\star$ tracks this transition, rising from $-0.888$ at the minimum target to 2.029 at the upper bound.  
Demanding target returns change the source of allocation sensitivity: concentration constraints become increasingly important in the upper frontier, where the portfolio may be compositionally fragile despite remaining mean-variance optimal.  

\begin{table}[htbp]
\centering
\caption{Dense-grid summary across selected target configurations.}
\label{tab:dense_summary}

\footnotesize
\setlength{\tabcolsep}{2.5pt}
\renewcommand{\arraystretch}{1.18}

\begin{tabular*}{\textwidth}{
@{\extracolsep{\fill}}
l
c c c c c c c
l
@{}
}
\toprule
Target configuration &
\(r^\star\) &
\(\sigma_p\) &
SR &
\(\overline{N}_{\mathrm{eff}}\) &
\(\overline{\mathrm{SD}}_w\) &
\(\mathrm{SD}_{w_{\mathrm{SPY}}}\) &
\(F_\star\) &
Driver \\
\midrule

Minimum target &
0.003 & 0.073 & 0.381 & 5.950 & 0.029 & 0.008 & -0.888 &
\(\ell_2\) penalty \\

Low target &
0.054 & 0.073 & 0.466 & 5.966 & 0.029 & 0.012 & -0.907 &
\(\ell_2\) penalty \\

Middle target &
0.079 & 0.080 & 0.747 & 5.686 & 0.030 & 0.025 & -0.694 &
\(\ell_2\) penalty \\

Driver shift &
0.111 & 0.100 & 0.910 & 4.387 & 0.037 & 0.043 & -0.009 &
weight cap \\

Upper target &
0.130 & 0.120 & 0.921 & 3.599 & 0.045 & 0.055 & 0.688 &
weight cap \\

Maximum target &
0.156 & 0.152 & 0.892 & 2.942 & 0.066 & 0.137 & 2.029 &
weight cap \\

\bottomrule
\end{tabular*}

\vspace{0.4em}

\parbox{0.98\textwidth}{
\footnotesize
\emph{Note:}
\(r^\star\) denotes the target return;
\(\sigma_p\) is portfolio volatility;
SR is the Sharpe ratio;
\(\overline{N}_{\mathrm{eff}}\) is the mean effective number of assets;
\(\overline{\mathrm{SD}}_w\) is the average standard deviation of portfolio weights;
and \(\mathrm{SD}_{w_{\mathrm{SPY}}}\) is the standard deviation of the SPY weight.
}

\normalsize
\end{table}

\subsection{Global Sensitivity Analysis and Parameter Interactions}
\begin{figure}[htbp]
    \centering
     \includegraphics[width=\textwidth]{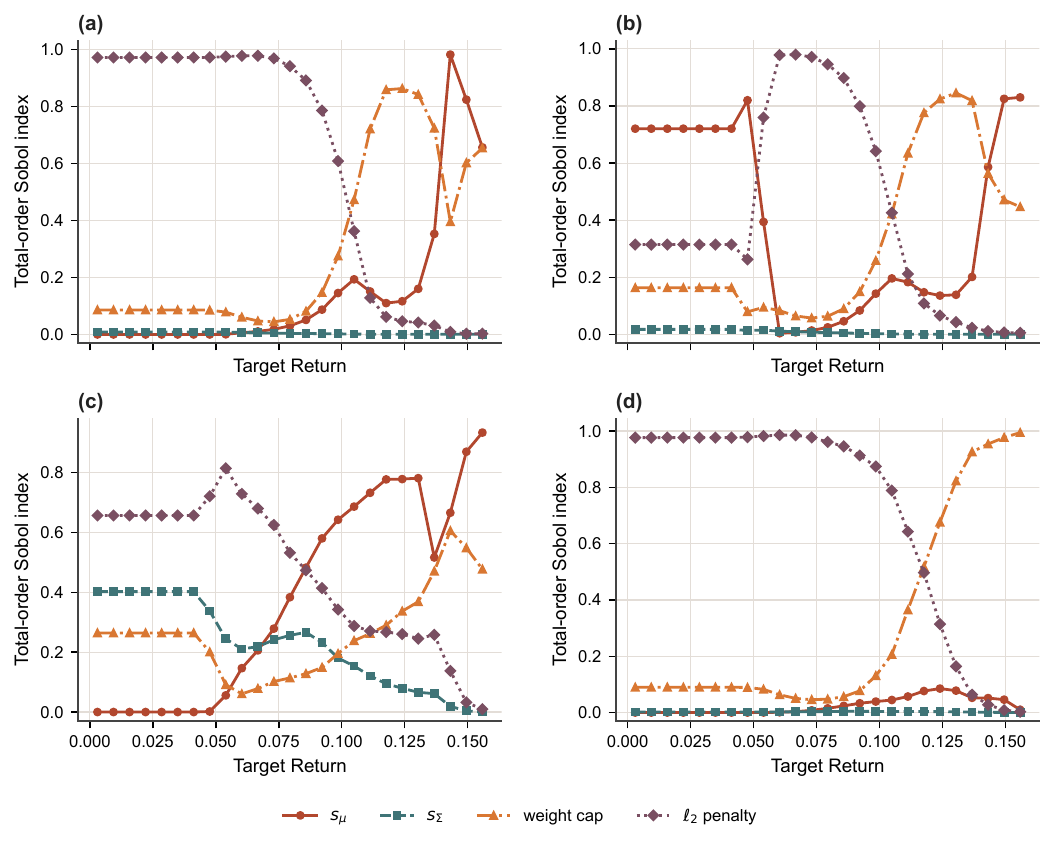}
    \caption{Total-order Sobol indices for
(a) portfolio variance,
(b) Sharpe ratio,
(c) SPY weight, and
(d) effective number of assets as functions of target return}
    \label{fig:sobolpaths}
\end{figure}

The sensitivity structure changes markedly across the target-return range. Figure~\ref{fig:sobolpaths} reports total-order indices across the target-return grid. Within the adopted stress domain, the lower and middle parts of the frontier are dominated by $\ell_2$ regularization, whereas the weight cap and expected-return perturbations gain importance as the return requirement becomes more demanding. Along the fixed covariance stress direction considered in the experiment, the covariance perturbation scale \(s_\Sigma\) remains comparatively weak.

Table~\ref{tab:s1_st_comparison} shows representative \(S_i\), \(S_{T_i}\), and interaction-gap values. For global outputs such as portfolio variance and effective diversification, $\ell_2$ regularization dominates in the lower and middle parts of the frontier, with moderate interaction gaps. By contrast, compositional outputs such as the weight of SPY show high interaction gaps even at low and middle target levels. Portfolio composition is therefore more strongly shaped by input interactions than aggregate risk or diversification measures.

\begin{table}[htbp]
\centering
\caption{Selected first-order and total-order Sobol indices across dense-grid target configurations.}
\label{tab:s1_st_comparison}
\footnotesize
\setlength{\tabcolsep}{3.5pt}
\renewcommand{\arraystretch}{1.10}

\begin{tabularx}{\textwidth}{
>{\raggedright\arraybackslash}p{1.55cm}
>{\raggedright\arraybackslash}p{1.85cm}
>{\raggedright\arraybackslash}p{1.55cm}
c c c
>{\raggedright\arraybackslash}X
}
\toprule
Output &
Target configuration &
Driver &
\(S_i\) &
\(S_{T_i}\) &
\(S_{T_i}-S_i\) &
Gap class \\
\midrule

Variance & Minimum target & \(\ell_2\) penalty &
0.918 & 0.973 & 0.055 & moderate \\

Variance & Middle target & \(\ell_2\) penalty &
0.913 & 0.942 & 0.029 & moderate \\

Variance & Driver-shift & weight cap &
0.705 & 0.721 & 0.016 & moderate \\

Variance & Maximum target & \(s_\mu\) &
0.332 & 0.657 & 0.324 & high \\

SPY weight & Minimum target & \(\ell_2\) penalty &
0.293 & 0.657 & 0.364 & high \\

SPY weight & Middle target & \(\ell_2\) penalty &
0.178 & 0.532 & 0.355 & high \\

SPY weight & Driver-shift & \(s_\mu\) &
0.434 & 0.733 & 0.299 & high \\

SPY weight & Maximum target & \(s_\mu\) &
0.520 & 0.934 & 0.414 & high \\

Effective \(N\) & Minimum target & \(\ell_2\) penalty &
0.918 & 0.977 & 0.058 & moderate \\

Effective \(N\) & Middle target & \(\ell_2\) penalty &
0.927 & 0.961 & 0.034 & moderate \\

Effective \(N\) & Driver-shift & \(\ell_2\) penalty &
0.554 & 0.643 & 0.088 & moderate \\

Effective \(N\) & Maximum target & weight cap &
0.990 & 0.994 & 0.004 & low \\
\bottomrule
\end{tabularx}

\vspace{0.35em}
\parbox{1\textwidth}{
\scriptsize\emph{Note:}
Gap classes are based on the empirical dense-grid baseline distribution
of \(S_{T_i}-S_i\), with \(q_{25}=0.004395\) and \(q_{75}=0.095772\).
}

\normalsize
\end{table}

The threshold diagnostics point to a progressive transition rather than a single abrupt threshold. The dense grid identifies a driver shift near 0.111 and a largest jump near 0.146, while the refined grid gives closely aligned values near 0.110 and 0.144. Together with the earlier onset proxy reported in Appendix~\ref{app:onset}, these diagnostics describe a sequence in which fragility first departs from its low-target regime, then changes sensitivity driver, and finally accelerates sharply in the upper tail. Around an annualized nominal target return of 0.11, the weight cap becomes a central determinant of optimizer behavior, while regularization loses relative influence. From roughly this point onward, concentration constraints become increasingly important for the stability of the allocation. The atlas next examines how this transition changes when the same concentration rule is applied to universes of different size.

\subsection{Fragility Atlas over Universe Size}
\begin{figure}[htbp]
    \centering
    \includegraphics[width=\textwidth]{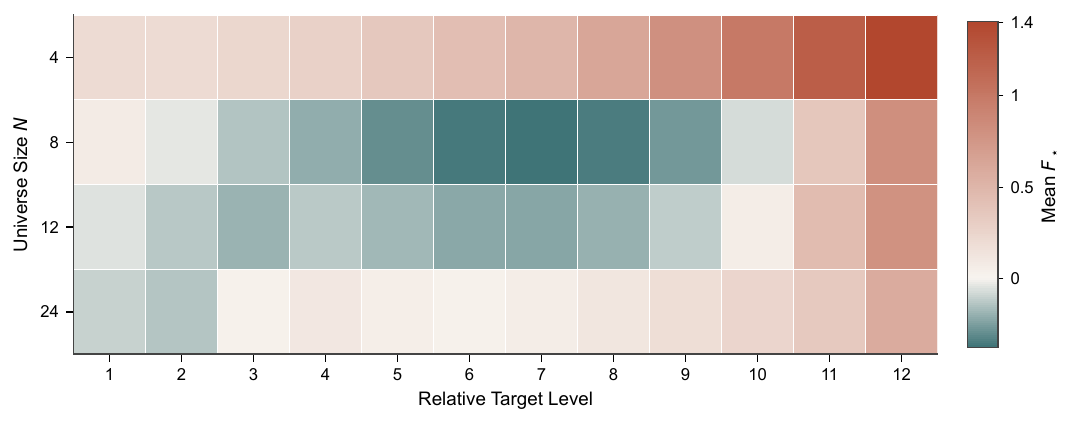}
    \caption{Mean $F_\star$ fragility score atlas over relative target-return levels and universe size $N$}
    \label{fig:atlas_fragility}
\end{figure}

\begin{figure}[htbp]
    \centering
    \includegraphics[width=\textwidth]{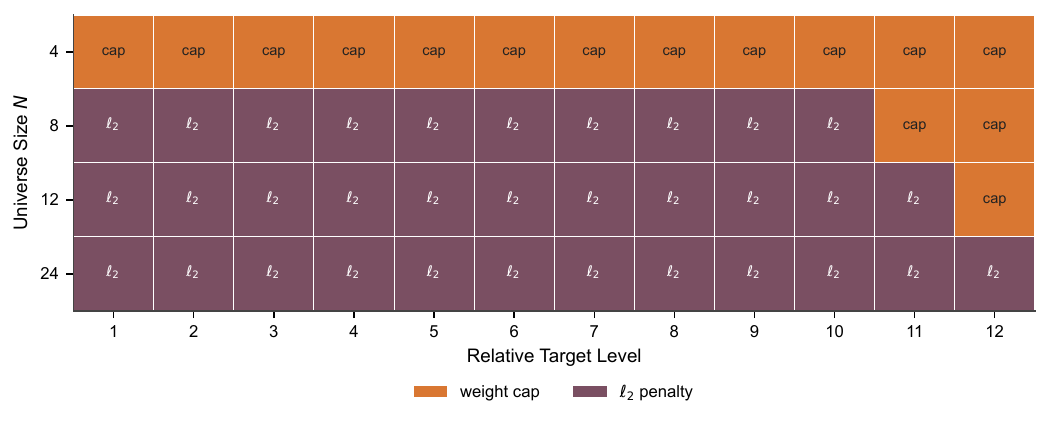}
    \caption{Dominant-driver atlas over relative target-return levels and universe size $N$}
    \label{fig:atlas_driver}
\end{figure}

Figure~\ref{fig:atlas_fragility} shows the mean $F_\star$ score over relative
target-return levels and $N$, while Figure~\ref{fig:atlas_driver} reports the
corresponding dominant-driver map across the same target levels and universe sizes.
Because the final atlas uses repeated nested-stratified sub-universe sampling from a
64-ETF master universe, the 24-asset case is now directly comparable to the smaller
universe sizes as a sampled sub-universe.

Table~\ref{tab:atlas_summary} summarizes the upper relative-target region. The smallest universe has the highest mean upper-region $F_\star$ and is the only case whose most frequent upper-region driver is the weight cap. For $N=8,12,24$, the most frequent driver remains the \(\ell_2\) penalty. This pattern reflects how a common absolute concentration rule interacts with the degrees of freedom available to the optimizer. With \(N=4\), fewer substitution possibilities make the weight cap a prominent driver in the aggressive-target region; larger universes can reallocate across a broader asset set, allowing regularization to remain the dominant stabilizing mechanism more often. For concentrated factor baskets or sector-rotation portfolios, concentration limits should therefore be calibrated with the size of the investable universe in mind.

\begin{table}[htbp]
\centering
\caption{Atlas summary by universe size in the upper relative-target region.}
\label{tab:atlas_summary}
\footnotesize
\setlength{\tabcolsep}{5pt}
\renewcommand{\arraystretch}{1.15}
\begin{tabularx}{\textwidth}{c c c c >{\raggedright\arraybackslash}X c c}
\toprule
$N$ &
\makecell{Mean\\Upper $F_\star$} &
\makecell{Std.\\Upper $F_\star$} &
\makecell{Mean Upper\\Eff. $N$ Ratio} &
\makecell{Most Frequent\\Upper Driver} &
\makecell{Mean Relative\\Onset Proxy} &
\makecell{Detected\\Replicates} \\
\midrule
4 & 1.009 & 0.904 & 0.819 & weight cap & 0.506 & 9/10 \\
8 & 0.097 & 0.437 & 0.514 & \(\ell_2\) penalty & 0.529 & 9/10 \\
12 & 0.202 & 0.403 & 0.399 & \(\ell_2\) penalty & 0.642 & 9/10 \\
24 & 0.297 & 0.208 & 0.231 & \(\ell_2\) penalty & 0.536 & 6/10 \\
\bottomrule
\end{tabularx}
\normalsize
\end{table}

The effective-number ratio declines with universe size in the upper region, from 0.819 for $N=4$ to 0.231 for $N=24$. 
Enlarging the universe does not therefore guarantee proportionally higher diversification: the optimizer tends to use a shrinking fraction of available assets as the target return rises and constraints become binding. 

\subsection{Robustness Checks}
The robustness checks are summarized here and reported in Appendix~\ref{app:robustness}. First, to assess whether the results depend on noise in the raw sample covariance matrix, the full baseline analysis is repeated after replacing the sample covariance with an RMT-filtered covariance matrix. The qualitative fragile-frontier pattern is preserved: low and middle target levels remain predominantly driven by \(\ell_2\) regularization, while the upper frontier remains governed by non-regularization drivers and exhibits higher weight dispersion. The exact dominant driver is preserved in 49 out of 50 covariance/grid/target comparisons; the only change occurs in one upper-tail refined target, where the dominant driver switches from the expected-return perturbation scale to the weight-cap constraint. RMT filtering also improves the conditioning of the covariance matrix, reducing the condition number from approximately 238.6 to 85.2.

Second, the correlated-input check with \(\rho=0.5\) between the expected-return and covariance perturbation scales preserves the qualitative ranking: based on absolute Pearson input-output associations, the low and middle target configurations remain most strongly associated with \(\ell_2\) regularization, while the high target remains associated with the weight-cap constraint. Third, the Sobol sample-size check over base sizes 32, 64, 128, 256, and 512 shows that the qualitative dominant-driver pattern is stable across all tested sizes. Across these checks, the broad low-to-high regime pattern remains unchanged under covariance filtering, moderate correlation between the market-stress channels, and different Sobol sampling budgets. 

\section{Discussion} 
\label{sec:discussion}
Portfolio stability changes systematically as return requirements become more demanding, pointing to a trade-off between return ambition and robustness to modeling and portfolio-construction inputs.
Two portfolios that appear close in the mean-variance plane may nevertheless differ substantially in their sensitivity to modeling inputs and portfolio-construction choices. 

First-order and total-order indices identify which inputs matter and whether their influence is direct or interaction-driven, a distinction that weight concentration alone cannot provide. Compositional instability can therefore be strongly interaction-driven even when aggregate risk measures appear comparatively stable. 

The atlas shows that this pattern also depends on universe size under a common concentration rule. What matters is not only the size of the universe, but how the number of available assets interacts with concentration limits and regularization as the return requirement becomes more demanding. In smaller universes the weight cap becomes more prominent, whereas broader asset menus provide greater scope for substitution and allow regularization to remain influential over a wider part of the target range. 

The analysis uses sample estimates of expected returns and covariances and a deliberately low-dimensional perturbation design based on fixed stress templates; as noted in Section~\ref{subsec:uncertainty_design}, the sensitivity rankings are tied to the adopted stress domain. The main Sobol analysis assumes independent inputs and is complemented by the correlated-input robustness exercise reported in Appendix~\ref{app:robustness}. Finally, $F_\star$ is used as the auxiliary standardized score defined in Section~\ref{subsec:fragility_score}. 

A natural next step is to examine whether the detected transition persists under rolling estimation, where parameter estimates and, potentially, the relevant perturbation domain evolve over time. A further theoretical extension is to formalize portfolio fragility in parametric optimization problems, where uncertain inputs induce a random optimal solution and fragility can be studied through the variability of decision-relevant functionals of the solution map. 

\section{Conclusion} 
\label{sec:conclusion}
We study how portfolio fragility changes as return requirements become more demanding and across investable universes of different size. The proposed diagnostic framework identifies where constrained Markowitz portfolios become more sensitive to jointly propagated uncertainty and portfolio-construction choices, and which inputs drive that sensitivity.

The baseline results show that portfolio fragility increases in the upper part of the target-return range. Within the adopted stress domain, the lower and middle regions are predominantly regularization-driven, while aggressive return requirements become increasingly sensitive to the weight cap and expected-return perturbations. Under the common concentration rule used in the atlas, sensitivity also varies systematically with universe size: the smallest-universe configurations are strongly weight-cap-driven, whereas larger sampled universes remain more often regularization-driven.

The fragile frontier therefore complements the efficient frontier by adding information on the stability of the optimized allocation. Moving upward along the frontier can mean moving not only toward higher risk, but also toward a region where the portfolio becomes more sensitive to modeling choices, uncertainty channels, and active constraints. Developing a formal characterization of this instability for broader classes of parametric optimization problems remains an open and non-trivial problem, since the solution map of a constrained optimizer is generally non-smooth in its inputs.

\backmatter

\bmhead{Statements and Declarations}

\bmhead{Funding}
Giulia Vannucci and Roberta Siciliano were supported by the Italian Ministry of Research, under the complementary actions to the NRRP ``Fit4MedRob -- Fit for Medical Robotics'' Grant (\# PNC0000007).

\bmhead{Competing Interests}
The authors have no relevant financial or non-financial interests to disclose.

\bmhead{Author Contributions}
Stefano Pellegrino developed the research idea, designed and implemented the methodology, conducted all empirical analyses, and wrote the original draft. Giulia Vannucci provided methodological guidance on the global sensitivity analysis framework, supervised the development of the Sobol-based diagnostics, and contributed to the revision of the manuscript. Roberta Siciliano provided overall supervision, contributed to the conceptual framing of the fragile-frontier diagnostic, and offered guidance on the empirical design and interpretation of the results. All authors reviewed and approved the final manuscript.

\bmhead{Data Availability}
Daily adjusted market-price data used in this study were obtained from Yahoo Finance for the period January 2018--December 2025. As these are third-party market data, the raw downloaded price files are not redistributed with the repository. The ticker definitions, data-retrieval and preprocessing code, portfolio configurations, derived analysis outputs, and reproducibility materials are publicly available in the accompanying GitHub repository and Zenodo archive.

\bmhead{Code Availability}
The Python code and reproducibility materials used to generate the analyses, tables, and figures reported in this study are publicly available at
\url{https://github.com/StefanoPellegrinoUnina/fragile-frontier-markowitz}
and are permanently archived on Zenodo
\citep{pellegrino2026fragilefrontiercode}
(version 1.0.0), under the MIT License.

\begin{appendices}

\section{Descriptive Figures}
\setcounter{figure}{4}
\renewcommand{\thefigure}{\arabic{figure}}
\setcounter{table}{5}
\renewcommand{\thetable}{\arabic{table}}

\begin{figure}[htbp]
    \centering
    \includegraphics[width=0.98\textwidth]{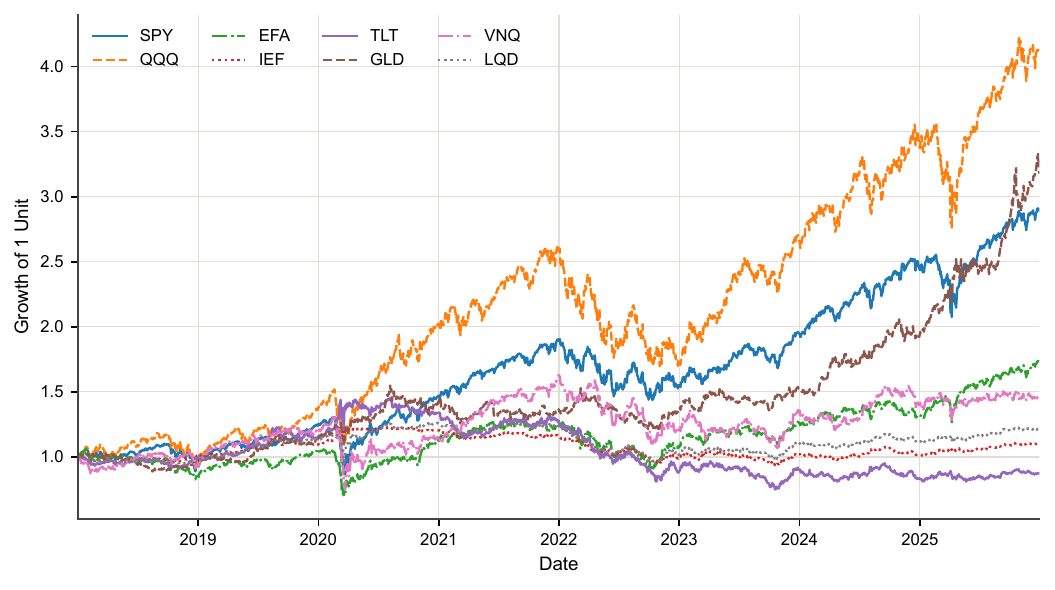}
    \caption{Normalized price paths for the baseline ETF universe}
    \label{fig:prices}
\end{figure}

\begin{figure}[htbp]
    \centering
    \includegraphics[width=0.7\textwidth]{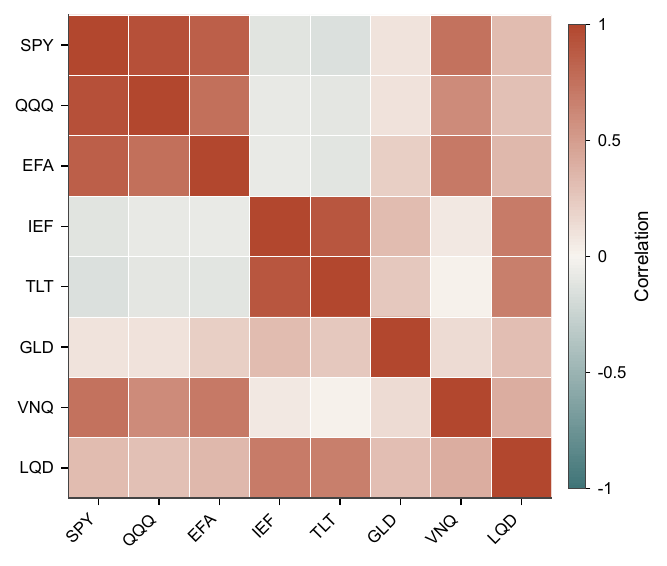}
    \caption{Return correlation matrix for the baseline ETF universe}
    \label{fig:corr}
\end{figure}

\label{app:descriptive}
Figures~\ref{fig:prices} and~\ref{fig:corr} provide descriptive views of the
normalized price evolution and return-correlation structure of the baseline ETF universe.

\section{Target-Grid Refinement and Onset Diagnostics}
\label{app:onset}

\begin{figure}[htbp]
    \centering
    \includegraphics[width=0.8\textwidth]{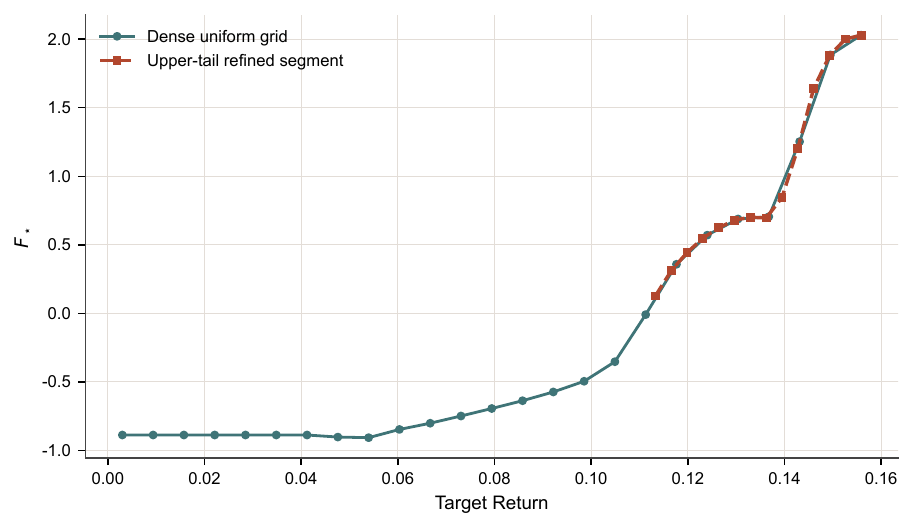}
    \caption{Dense-grid $F_\star$ curve and upper-tail refined segment}
    \label{fig:refined_grid}
\end{figure}

\begin{figure}[htbp]
    \centering
    \includegraphics[width=\textwidth]{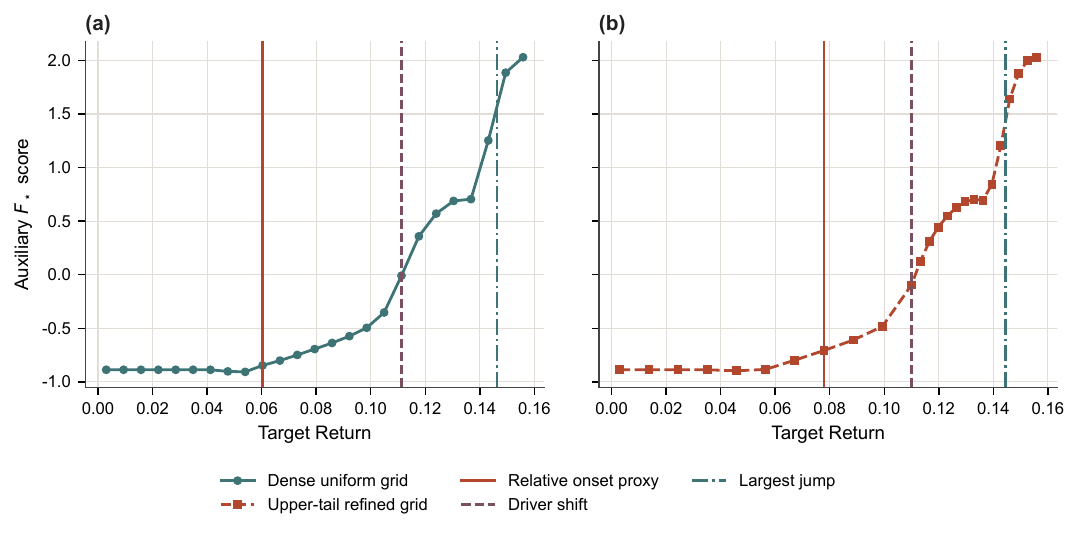}
    \caption{Comparison of relative onset, driver-shift, and largest-jump diagnostics}
    \label{fig:threshold}
\end{figure}

Figure~\ref{fig:refined_grid} shows the upper-tail refinement of the baseline
fragility curve. Table~\ref{tab:threshold_robustness} reports the corresponding
onset diagnostics across grid designs, while Figure~\ref{fig:threshold} provides
a graphical comparison of the relative-onset, driver-shift, and largest-jump indicators.

\begin{table}[htbp]
\centering
\caption{Onset and transition diagnostics across grid designs using the final $F_\star$ score.}
\label{tab:threshold_robustness}

\footnotesize
\setlength{\tabcolsep}{5pt}
\renewcommand{\arraystretch}{1.15}

\begin{tabular}{l c c c c}
\toprule
Grid design &
\makecell{Relative onset\\proxy} &
\makecell{Driver\\shift} &
\makecell{Largest\\jump} &
\makecell{Stable across\\refinements} \\
\midrule

Dense uniform &
0.060 & 0.111 & 0.146 & True \\

Upper-tail refined &
0.078 & 0.110 & 0.144 & True \\

\bottomrule
\end{tabular}

\normalsize
\end{table}

\section{Master Universe and Basket Design}
\label{app:master_universe}

Table~\ref{tab:master_universe} reports the complete 64-ETF master universe used
for the repeated nested-stratified atlas and identifies the ETFs included in the
baseline universe.

\begin{table}[htbp]
\centering
\caption{Final 64-ETF master universe used for the repeated nested-stratified atlas. Baseline ETFs are shown in bold.}
\label{tab:master_universe}
\scriptsize
\setlength{\tabcolsep}{4pt}
\renewcommand{\arraystretch}{1.12}
\begin{tabularx}{\textwidth}{l c X}
\toprule
Stratum & Count & Tickers \\
\midrule
Emerging Market Equity & 10 & EEM, IEMG, VWO, SCHE, EWZ, FXI, INDA, EWT, EWY, EWW \\
Fixed Income & 12 & SHY, \textbf{IEF}, \textbf{TLT}, \textbf{LQD}, HYG, BND, AGG, TIP, MUB, EMB, BIL, VCIT \\
International Developed Equity & 10 & \textbf{EFA}, VEA, VXUS, IEFA, EFV, EFG, EWJ, EWU, EWC, EWL \\
Real Assets & 10 & \textbf{GLD}, IAU, SLV, \textbf{VNQ}, DBC, USO, DBA, PDBC, RWO, REET \\
U.S. Broad/Style/Factor & 12 & \textbf{SPY}, VTI, \textbf{QQQ}, IWM, DIA, RSP, IVV, VOO, MTUM, QUAL, USMV, VLUE \\
U.S. Sector Equity & 10 & XLB, XLE, XLF, XLI, XLK, XLP, XLU, XLV, XLY, SMH \\
\bottomrule
\end{tabularx}
\normalsize
\end{table}

\section{Additional Robustness Checks}
\label{app:robustness}

This appendix reports three robustness checks supporting the Sobol-based interpretation in the main text. The first check evaluates whether the baseline fragile-frontier pattern depends on noise in the raw sample covariance matrix by repeating the full baseline analysis with an RMT-filtered covariance matrix. The second check introduces moderate positive dependence between the expected-return and covariance perturbation channels and examines whether the qualitative association pattern is preserved. The third check evaluates whether the dominant-driver pattern is stable as the Sobol base sample size increases toward the final reference value \(N_S=512\).

\subsection{RMT-filtered covariance check}

To evaluate whether the baseline results are driven by noise in the sample covariance matrix, the full baseline analysis is repeated after replacing the raw sample covariance matrix with an RMT-filtered covariance estimate. The expected-return vector \(\hat{\mu}\), the target grids, the Sobol input space, the perturbation design, and the constrained Markowitz formulation are kept unchanged. The check is therefore a covariance-filtering robustness exercise rather than a separate portfolio model.

The filter is applied to the sample correlation matrix. Letting \(q=N/T\), where \(N\) is the number of assets and \(T\) the number of return observations, the Marchenko--Pastur upper edge is
\[
\lambda_{+}=(1+\sqrt{q})^2.
\]
In this implementation, eigenvalues not exceeding \(\lambda_{+}\) are treated as belonging to the noise bulk and replaced by their average, while the corresponding eigenvectors are left unchanged. The filtered correlation matrix is then renormalized and rescaled using the original asset volatilities to recover the covariance matrix. The resulting covariance matrix is finally symmetrized and its eigenvalues are floored at \(10^{-8}\), consistently with the covariance treatment used in the main analysis.

The qualitative regime structure of the fragile frontier is preserved. Across the dense and upper-tail refined grids, the exact dominant driver is preserved in 49 out of 50 covariance-grid-target comparisons. The only exact driver change occurs in one upper-tail refined configuration, where the dominant driver switches from \(s_\mu\) to weight cap. This change occurs in the upper frontier and does not alter the main interpretation: the upper target region remains non-regularization-driven and more weight-dispersed than the lower region.

RMT filtering also improves numerical conditioning, reducing the covariance condition number from approximately 238.6 to 85.2. The main driver-shift and largest-jump diagnostics remain aligned with the original baseline. The relative-onset proxy moves from an earlier signal under the sample covariance to approximately the same level as the driver-shift threshold under RMT filtering. This suggests that covariance filtering attenuates early fragility signals while preserving the main transition region of the fragile frontier.

\subsection{Correlated-input check}

In the correlated-input check, \(s_\mu\) and \(s_\Sigma\) are coupled through a Gaussian copula with \(\rho=0.5\) at representative low, middle, and high target-return levels. Absolute Pearson correlations between inputs and portfolio outputs are used to summarize the corresponding association pattern. The qualitative ranking is unchanged: low and middle targets remain most strongly associated with the \(\ell_2\) penalty, whereas the high target remains associated with the weight cap. Under this moderate dependence, the qualitative low/middle/high input-output association pattern remains unchanged.

\subsection{Sobol sample-size convergence}

\begin{figure}[!htbp]
    \centering
    \includegraphics[width=0.85\textwidth]{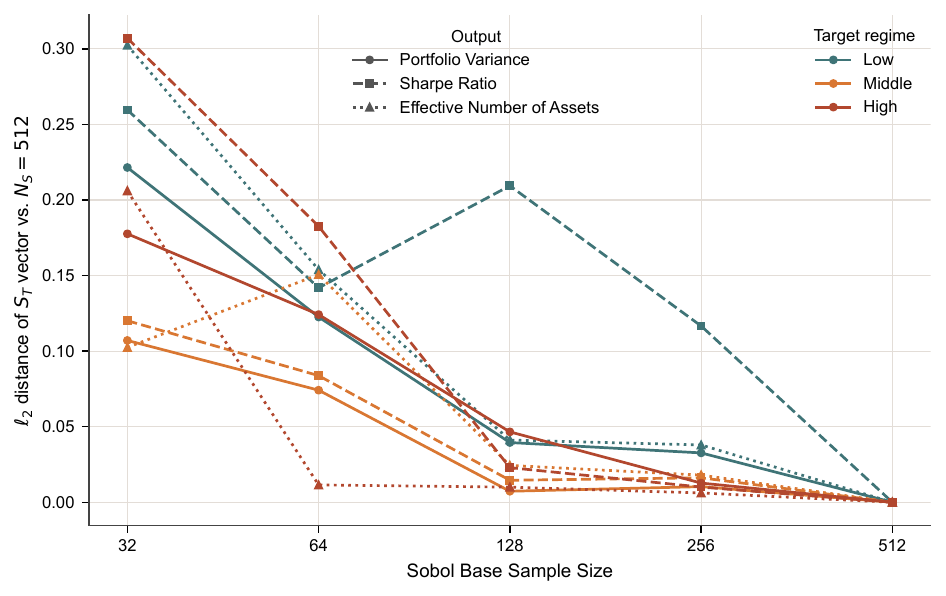}
    \caption{Practical Sobol sample-size convergence diagnostic toward the \(N_S=512\) reference}
    \label{fig:sobol_sample_size_convergence}
\end{figure}

Figure~\ref{fig:sobol_sample_size_convergence} reports the distance between total-order sensitivity vectors computed at smaller Sobol base sample sizes and the final \(N_S=512\) reference. The distances generally decrease as the sample size increases, and the qualitative dominant-driver ranking remains stable across \(N_S \in \{32,64,128,256,512\}\). In all tested designs, low and middle target configurations remain regularization-driven, whereas the high target configuration remains weight cap-driven.

\end{appendices}

\bibliography{references}

\end{document}